\documentclass[a4paper]{article}

\usepackage{icrc2013}
\usepackage[english]{babel}
\usepackage{amssymb}

\title{Dark Matter 2013}

\shorttitle{Dark Matter 2013}

\authors{
Marc Schumann.
}

\afiliations{
Albert Einstein Center for Fundamental Physics, University of Bern, Switzerland \\
}

\email{marc.schumann@lhep.unibe.ch}

\abstract{This article reviews the status of the exciting and fastly evolving field of 
dark matter research as of summer 2013, when it was discussed at ICRC~2013 in Rio de Janeiro.
It focuses on the three main avenues to detect WIMP dark matter: direct detection, indirect detection and collider searches. 
The article is based on the dark matter rapporteur talk summarizing the presentations given at
the conference, filling some gaps for completeness.}

\keywords{ICRC2013, dark matter, direct detection, indirect detection, collider searches, WIMP.}

\begin{document}
\maketitle

\section{Introduction}

\begin{figure}[h!]
 \centering
 \includegraphics[width=0.4\textwidth]{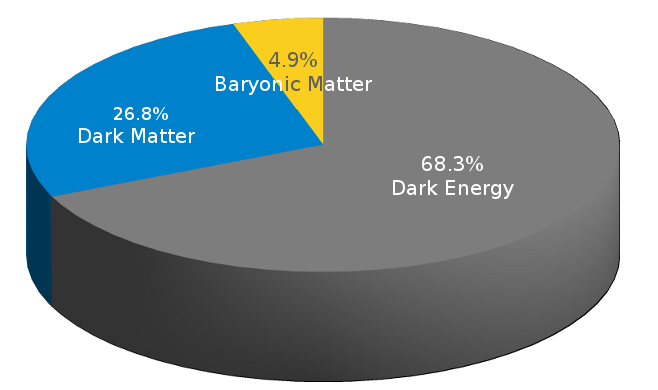}
 \caption{The latest results from the Planck satellite~\cite{ref::planck} on the energy densities attributed to dark energy, dark matter and 'ordinary' baryonic matter. }
 \label{fig::planck}
\end{figure}

The existence of dark matter is one of the stongest indications that there must be physics beyond the standard model of particle physics. Numerous indirect observations at astronomical and cosmological scales~\cite{ref::evidence}, complemented by results of complex many-body simulations~\cite{ref::simulation}, point to the presence of a new form of matter in the Universe, which only interacts significantly via gravity. The most famous observational evidence is the rotation profiles of galaxies, the dynamics of galaxy clusters, the separation of dark and light matter in galaxy clusters, and the interpretation of the cosmic microwave background (CMB). Recently the Planck satellite mission~\cite{ref::planck} has published new and precise measurements of the CMB, which are in full agreement with the predictions of the $\Lambda$CDM model, describing a cosmos dominated by dark energy ($\Lambda$) and cold dark matter (CDM). The new values for the energy densities from Planck are $\Omega_{CDM}=0.268$ and $\Omega_\Lambda=0.683$, see Fig.~\ref{fig::planck}.

Even though dark matter makes up a sizeable fraction of the energy density of the Universe, and outnumbers 'ordinary' baryonic matter by a factor~5, the particle(s) which constitute the dark matter remain unknown as of today. The absence of electromagnetic and strong interactions makes it experimentally 'dark', however, interactions at the weak scale might be possible. Many theories beyond the standard model predict particles which are neutral, cold (i.e.~non-relativistic), and stable (or have half-lifes longer than the age of the Universe). These are viable dark matter candidates, with the most prominent being the neutralino $\chi_0$ in supersymmetric theories~\cite{ref::susy}, the lightest Kaliza-Klein particle (LKP) in theories with extra-dimensions~\cite{ref::extradim}, or the lightest $T$-odd particle in little Higgs models~\cite{ref::littlehiggs}. All are excellent examples of weakly interacting massive particles (WIMPs)~\cite{ref::wimps}, which are stable as their decays are prevented by some new symmetry.

The search for the dark matter particle has become one of the most exciting topics in Astroparticle Physics, and tremendous progress is made on experimental and theoretical research. For this reason, this year's ICRC conference in Rio de Janeiro (``The Astroparticle Conference'') featured, for the first time, a full branch dedicated to dark matter. This article is the attempt to summarize the main conclusions of the talks and posters presented at this occasion, complemented with some extra information added by the author. At this occasion, the author wants to apologize to all contributors to the ICRC 2013 dark matter session, whose work could not be mentioned in this highly-biased summary.

The article contains three sections addressing the different methods to detect WIMP dark matter: by searching for signs of WIMPs scattering in low-background detectors (direct detection, Sect.~\ref{sec::direct}), by looking for WIMP annihilation products (indirect detection, Sect.~\ref{sec::indirect}), and by searching for WIMPs produced in particle colliders such as the LHC (Sect.~\ref{sec::collider}). The approaches are largely complementary and it is widely assumed that a convincing dark matter signal should be seen by more than one. We do not even attempt to provide detailed descriptions of the various experiments, but mainly focus on the underlying concepts and the recent results, and refer the reader to the references for further information. The article closes with a short section on more exotic (here: ``non-WIMP'') dark matter models and a conclusion.

\section{Direct Detection}\label{sec::direct}

It has been pointed out by Goodman and Witten~\cite{ref::directdet} in 1985, that the signature of WIMPs scattering in a detector medium might be directly detectable by sensitive instruments~\cite{ref::baudis}, provided that the WIMP interacts not only gravitationally with ordinary matter but with weak-scale cross sections. Another prerequisite is that there is dark matter in our local solar neighborhood, which is assumed to be the case as confirmed by various astronomical studies~\cite{ref::localdm}. The canonical value for the local WIMP density used for the interpretation of measurements is $\rho_{DM}=0.3$\,GeV/cm$^3$.

Being neutral particles and moving at non-relativistic velocities, WIMPs are expected to interact mainly with the atomic nucleus, whose nuclear recoil energy is to be measured by the dark matter detector. The recoil spectrum is a featureless falling exponential, only modified by form factor corrections for heavier nuclei, and kinematics fixes the maximum energies at a few tens of keV. As it is a priori not known how WIMPs interact with the detector matter, two cases are typically considered. The first one is a spin-independent (SI) scalar interaction with the WIMP-nucleus cross section given by
  $$\sigma_{SI}=\sigma_n \frac{\mu_N^2}{\mu_n^2}\frac{(f_pZ+f_n(A-Z))^2}{f_n^2} = \sigma_n \frac{\mu_N^2}{\mu_n^2} A^2 \textnormal{,}$$
where the last equality assumes that the WIMP couplings $f_{p,n}$ to protons and neutrons are identical, leading to an $A^2$ dependence of the cross section. $\sigma_n$ is the scattering cross section on a nucleon (making comparisons between different targets easier), and $\mu_{n,N}$ are the reduced masses of the WIMP-nucleon and WIMP-nucleus systems, respectively. In the second case, spin-dependent (SD) axial vector couplings, the differential WIMP-nucleus cross section depends on the momentum transfer $\vec{q}$ and reads 
  $$\frac{d\sigma_{SD}}{\textnormal{d}|\vec{q}|^2} = \frac{8 G_F^2}{\pi v^2} \left[ a_p \langle S_p \rangle + a_n \langle S_n \rangle \right]^2 \frac{J+1}{J} \frac{S(|\vec{q}|)}{S(0)} \textnormal{.} $$
$S(|\vec{q}|)$ is the spin-structure function, $a_{p,n}$ the couplings to protons and neutrons, and $\langle S_{p,n} \rangle$ are the expectation values of the total spin operators in the nucleus. In this case, no $A^2$ enhancement is present but the sensitivity depends crucially on the spin-structure of the particular target nucleus and on its total nuclear spin $J$, leading to a very different picture compared to the spin-independent case. For simplicity, spin-dependent results are usually reported assuming that WIMP couple to protons ($a_n=0$) or neutrons only ($a_p=0$).

The dominating backgrounds in WIMP searches are $\gamma$-rays from the environment or the experimental setup itself and $\beta$-particles at the surfaces or in the bulk of the detector. They interact electromagnetically with the atomic electrons, leading to electronic recoils (ER)	. The different ionization density of electronic and nuclear recoil (NR) interactions is often used to discriminate signal (NR) from background (ER). $\alpha$-contamination in the detector materials does usually not pose a problem when the full energy is detected, they might become relevant when lots of the $\alpha$-energy is lost in insensitive detector regions. 

However, the most dangerous background for all direct WIMP searches are neutron-induced nuclear recoil interactions since these cannot be distinguished from a WIMP signal. The only important difference is the event multiplicity: the WIMP-nucleus interaction cross section is expected to be extremely small, which means that WIMPs will always scatter only once in a detector. Neutrons, on the other hand, will often produce double-scatter signatures. An effective WIMP detector must therefore be able to identify (and reject) events with multiple interactions. In many detectors, the background can be further reduced by exploiting the self-shielding capability of the target material, which is especially effective for high-$Z$ materials, by rejecting events occuring in the outer detector regions (``fiducialization''). This requires that the position of the interaction vertex can be measured with some precision.

Most of the backgrounds can be effectively suppressed by means of massive shields surrounding the detectors, either being made of high-$Z$ materials such as lead and copper, plus polyethylene to reduce the neutron flux, or consisting of several meters of water. However, the only way to reduce the background due to muon-induced neutrons to acceptable levels is to reduce the muon flux by several orders of magnitude compared to the one at sea level. Therefore, all dark matter detectors are placed in deep underground laboratories, with typically 1-2\,km of rock overburden, suppressing the muon flux by 5-7\,orders of magnitude.

\begin{figure}[t]
 \centering
 \includegraphics[width=0.5\textwidth]{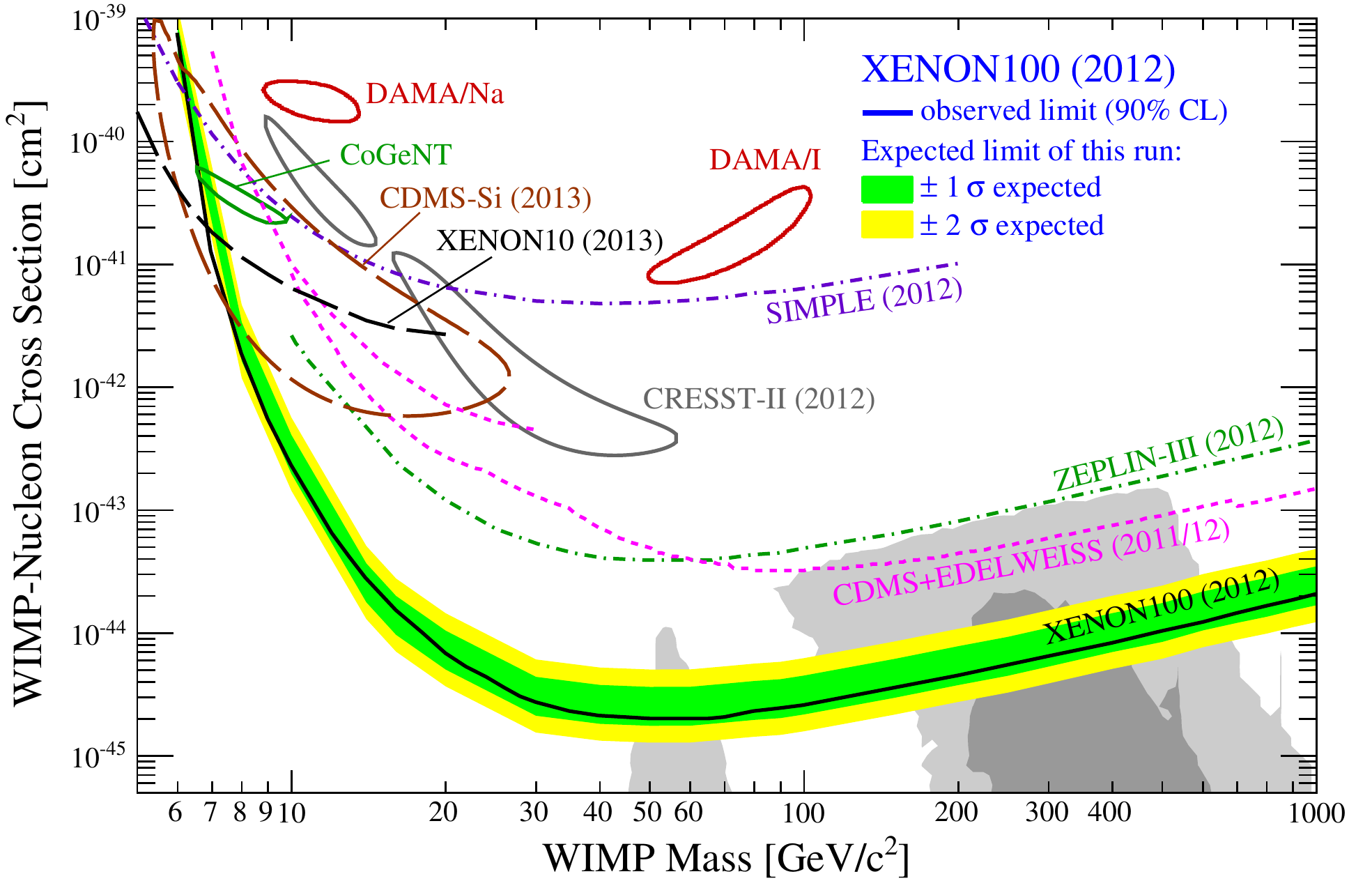}
 \caption{The current experimental results on spin-independent WIMP-nucleon scattering cross sections can be divided into a low ($\le$30\,GeV/$c^2$) and a high-mass region. In the first one, several hints for WIMPs seen by CoGeNT~\cite{ref::cogent}, CRESST-II~\cite{ref::cresst}, DAMA/Libra~\cite{ref::dama,ref::damawimp}, and CDMS-Si~\cite{ref::cdms_si} (the $2\sigma$ regions are shown) are challenged by the null-results from XENON100~\cite{ref::xe100}, XENON10~\cite{ref::xe10_corr}, EDELWEISS~\cite{ref::edelweiss,ref::edelweiss_low} and ZEPLIN-III~\cite{ref::zeplin}. The high-mass region is dominated by the results of XENON100, due to its large exposure and low background. Fig.~adapted from~\cite{ref::xe100}.}
 \label{fig::wimps}
\end{figure}

The current parameter space for spin-independent scattering cross sections obtained from direct WIMP searches is shown in Fig.~\ref{fig::wimps}. It can be divided into a high-mass region above WIMP masses of $\sim$30\,GeV/$c^2$, and a low-mass region below. Especially the latter is of special interest at the moment, as hints for WIMP signals around masses of 10\,GeV/$c^2$ from several experiments are confronted with null-results from others.

The DAMA/Libra experiment~\cite{ref::dama} looks for an annually modulating recoil spectrum as a consequence of the Earth's movement around the Sun during the year, making the expected recoil spectrum harder or weaker depending on whether the Earth's velocity is added or subtracted to the Sun's velocity around the Galactic center~\cite{ref::annualmod}. The DAMA collaboration employs a large mass of ultra-pure NaI(Tl) scintillators and observes such a modulation signal with a high significance since many years. However, it appears to be rather incompatible with other results when it is interpreted as being due to WIMP interactions~\cite{ref::damawimp}. 

The CoGeNT experiment is based on a rather low-mass ($\sim$330\,g only) p-type Ge-detector, which features a very low energy threshold. A first analysis reporting an unexpected exponential-shaped excess of events at very low energies~\cite{ref:cogent_old} was recently updated using a new analysis: the size of the possible signal decreased because of an improved way to reject surface events, shifting the previously larger signal region to lower cross-sections~\cite{ref::cogent}. In addition, CoGeNT also reported an annually modulating signal~\cite{ref::cogent_mod}, which, however, shows a rather unexpected energy behavior and appears to be in conflict with the non-observation of a modulation signal in a dedicated analysis by the CDMS-II collaboration, which also uses Ge~detectors~\cite{ref::cdms-mod}. Another excess of events was observed by the CRESST experiment~\cite{ref::cresst} detecting light and heat signals from CaWO$_4$ crystals cooled to mK temperatures. While DAMA and CoGeNT only measure one observable (light or ionization, respectively), the two observables detected by CRESST can be used to discriminate signal (NR) from background (ER). Still, a total of 67\,events was observed in a WIMP search region, where the excess above the background expectation of $\sim$40\,events could be explained by a WIMP signal. 

In April 2013, the CDMS collaboration, operating Ge detectors at cryogenic temperatures to measure the ionization and heat signal from particle interactions~\cite{ref::cdms_si}, and which are capable of fiducialization, reported a new result from the Si~detectors which are installed in the same setup in the Soudan laboratory, USA. Three events were observed, close the threshold as expected from WIMP dark matter, while only 0.41~events were expected from the background model. This excess can be interpreted as being due to low-mass WIMP interactions, with the point of highest likelihood being at a mass of 8.6\,GeV/$c^2$ and a cross section of $1.9\times 10^{-41}$\,cm$^2$. 

The question remains whether all these positive hints are of common origin, maybe from WIMP interactions, or whether they are due to non-understood backgrounds. At face value, the signal regions do not all overlap, however, one can find explanations which brings them to agreement. The situation is even more complicated as there are several other experiments which do not see an excess. One of them is CDMSlite, a 600\,g germanium detector operated in a mode where the ionization signal is Neganov-Luke amplified. A very low threshold of 0.17\,keV was reached and a 10\,live-days exposure was sufficient to exclude the upper part of the CDMS-Si region~\cite{ref::cdmslite}.

Basically all positive signals are challenged by the null-result of the XENON100 collaboration, which operates a dual-phase time projection chamber (TPC) at LNGS, Italy. The TPC is filled with ultra-pure liquid xenon (LXe) at a temperature of $-93^\circ$C, in order to detect light and charge induced by particle interactions. Thanks to its large target mass of 62\,kg and a background level which is superior to all other current experiments~\cite{ref::xe100_em}, the collaboration was able to report a very strong exclusion limit after running for 225\,live days and observing no significant excess of events in an inner 34\,kg fiducial LXe target~\cite{ref::xe100}. Only a small fraction of the CDMS-Si region is not excluded (at 90\% CL) by XENON100 and an older XENON10 result, which was corrected recently~\cite{ref::xe10_corr}. Noble gas TPCs feature a lower energy resolution than crystal detectors (Ge, NaI), and the signal quenching of NRs and therefore the energy scale is less well known. However, the XENON collaboration demonstrated in a recent publication that the detector response is very well understood down to nuclear recoil energies of 3\,keV~\cite{ref::xe100_mcmatch}.

The entire CRESST region, as well as most part of the DAMA and CDMS-Si regions, are excluded by a dedicated low-mass analysis of EDELWEISS-II~\cite{ref::edelweiss, ref::kleifges}. This experiment installed in the LSM laboratory in France also uses cryogenic Ge detetors, measuring charge and heat signals, and reached a high trigger efficiency of $>$75\% at 5\,keV in this analysis. The upcoming phase EDELWEISS-III aims at achieving competitive sensitivities around $10^{-45}$\,cm$^2$ by 2014/15. It will use improved Ge detectors of 800\,g mass, out of which $\sim$600\,g will remain after fiducialization and which also allow for better background rejection.

A different approach than the ones described so far is pursued by XMASS. This Japanese experiment uses a massive target of 835\,kg of LXe, however, not operated in a TPC but as a single-phase detector in which only the light signal is detected~\cite{ref::yamashita}. Background rejection is entirely based on fiducialization as no other means of discrimination exist. Do to its single-phase nature and an optimized design, the detector features a very high light yield and a first analysis using the full target cuts into the DAMA preferred region~\cite{ref::xmass}. The next stage of the experiment aims at reducing the background by refurbishing the detector and has started data taking this summer~\cite{ref::martens}. 

Because of their finite energy threshold, all current WIMP experiments show a strong decrease in sensitivity below $\sim$10\,GeV/$c^2$. The sensitivity-loss is more pronounced for heavier targets such as xenon, whereas the p-type Ge detector experiments CoGeNT~\cite{ref::cogent}, TEXONO~\cite{ref::texono} and CDEX~\cite{ref::cdex} have published results down to WIMP masses of 4\,GeV/$c^2$. A new experiment, DAMIC, aims at pushing down the sensitivity to $\sim$1.5\,GeV/$c^2$ by using low-noise, low-threshold CCD chips as WIMP target to measure the ionization energy deposited in the Si. A first result~\cite{ref::damic,ref::tiffenberg} sets the best limit below 4\,GeV/$c^2$, however, still at rather high cross sections of a few $10^{-39}$\,cm$^2$. A new run with the new experiment DAMIC100 will improve this result by a factor~100 after an exposure of 1\,y$\times$100\,g .

\begin{figure}[h!]
 \centering
 \includegraphics[width=0.5\textwidth]{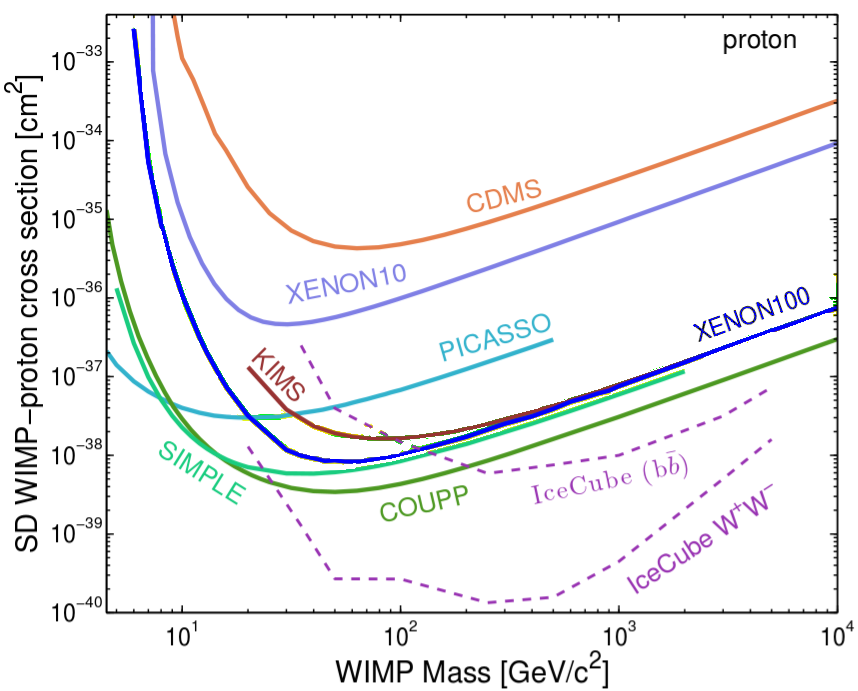}
 \includegraphics[width=0.5\textwidth]{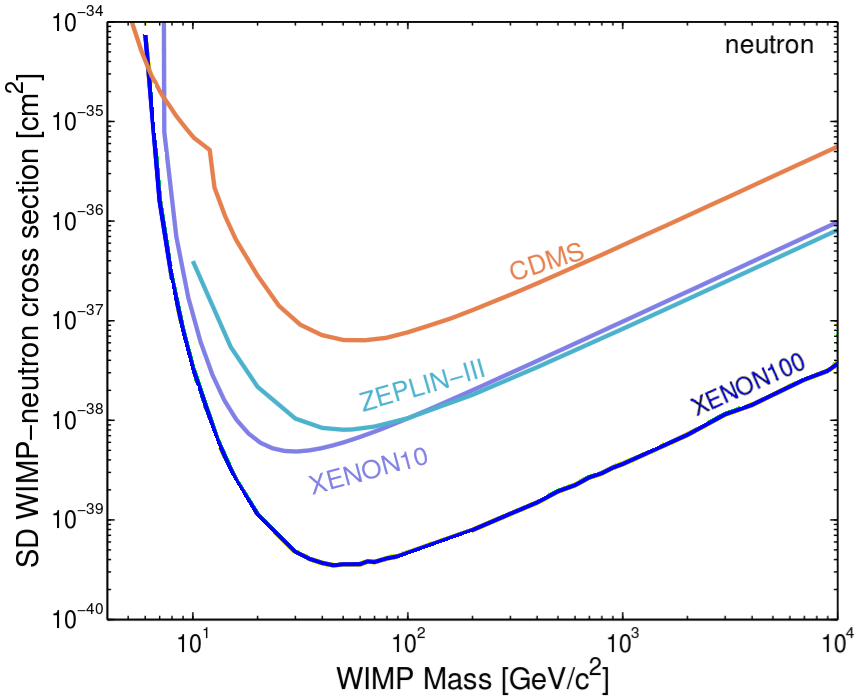}
 \caption{Results from direct detection experiments on spin-dependent WIMP-nucleon cross sections are usually separated into proton-only and neutron-only couplings. In the first case (top), the parameter space is dominated by results from experiments using a target with $^{19}$F (PICASSO~\cite{ref::picasso, ref::noble}, SIMPLE~\cite{ref::simple}, COUPP~\cite{ref::coupp}). Indirect searches (see Sect.~\ref{sec::indirect}) conducted by experiments such as IceCube~\cite{ref::icecube_sd} also provide very strong limits for this channel. The best limit on  neutron-only couplings (bottom) is from XENON100~\cite{ref::xe100_sd}. Figs.~adapted from~\cite{ref::xe100_sd}, further references to the shown results can be found there.}
 \label{fig::sd}
\end{figure}

The discussion so far was entirely focused on spin-independent WIMP nucleon couplings. As mentioned above, the sensitivity to spin-dependent cross sections does not scale with $A^2$ but depends on the unpaired nuclear spins $J$ and the nuclear spin expectation values. The isotope with the highest spin enhancement factor $(J+1)/J \times \left[ a_p \langle S_p \rangle + a_n \langle S_n \rangle \right]^2$ is $^{19}$F.  Therefore it is the most sensitive target for spin-dependent interactions and is being used in bubble chambers and superheated droplet detectors. An example is PICASSO~\cite{ref::noble}, which consists of small droplets of C$_4$F$_{10}$ embedded in a gel. The droplets are superheated, i.e., they are maintained at a temperature higher than their boiling point. By tuning the operating conditions, these detectors can be made insensitive to $\gamma$- and $\beta$-backgrounds, while nuclear recoil interactions (from WIMPs or neutrons) and $\alpha$s inject enough energy into the bubbles to cause a phase transition. These detectors are rare examples where $\alpha$-particles pose a serious background, however, the acoustic signal of the bubble formation can be used to distinguish them from NRs. The specific spin-structure of $^{19}$F makes it very sensitive to proton-only couplings ($a_n=0$): the experiments PICASSO~\cite{ref::picasso}, SIMPLE~\cite{ref::simple} and COUPP~\cite{ref::coupp} place the most stringents limits as no excess of events above background has been observed in any of the instruments, see Fig.~\ref{fig::sd} (top). In the neutron-only case ($a_p=0$), the highest sensitivity was achieved by XENON100~\cite{ref::xe100_sd} which sets the best upper limit over the full mass range. The XENON100 constraint to proton-only couplings is almost 2~orders of magnitude weaker, underlining again the fact that the spin-dependent sensitivity depends on the details of the nuclear structure. 

There are more experiments searching for WIMPs, which have not been addressed in this overview because they did not publish science results yet, such as LUX\footnote{LUX has announced very strong constraints on spin-independent couplings after the completion of this review~\cite{ref::lux}.}, DarkSide, DEAP, CLEAN, PandaX, etc. Additionally, most experiments mentioned so far plan to increase their target mass in the near future. PICASSO and COUPP, for example, have merged to the PICO collaboration which aims for bubble chamber with $\sim$500\,kg target mass in the next 2\,years~\cite{ref::noble}. CDMS-II is moving towards SuperCDMS, which is likely to be installed at SNOLAB, with a planned Ge target of about 100\,kg. The XENON collaboration has started the underground construction of XENON1T at LNGS this summer. The new dual-phase TPC will feature a total mass of $\sim3000$\,kg out of which more than 1000\,kg will be used as fiducial target. Detector operation will start in 2015 with the goal to reach spin-independent cross sections of $2\times10^{-47}$\,cm$^2$ after 2\,years of data taking. This is two orders of magnitude beyond than the current best results. In the longer term, the collaboration will upgrade XENON1T to the even larger version XENONnT with 3-4\,t of fiducial mass. It will be able to reach sensitivities at which neutrino interactions constitute a non-negligible background~\cite{ref::neutrinos}.

\section{Indirect Detection}\label{sec::indirect}

While direct detection experiments discussed in the previous section require that there is dark matter present in the Earth's neighborhood, this is not generally necessary for indirect detection approaches. These assume that WIMPs are their own anti-particles (as it is the case for Majorana particles such as the SUSY neutralino) which will annihilate in standard model particles if they collide because their local density is large enough. Possible annihilation scenarios include 
  $$\chi \overline{\chi} \ \to \ q \overline{q}\textnormal{,} \ \ell \overline{\ell} \textnormal{,} \ W^+W^-\textnormal{,} \ ZZ \textnormal{.}$$
These primary particles eventually decay into positrons, electrons, anti-protons, protons, neutrinos and $\gamma$-rays, which can be observed by dedicated instruments. As it is generally not known which particles are preferentially generated in the WIMP annihilation, various channels are usually considered in the analysis independently ($\chi \overline{\chi} \to b \overline{b}$, $\chi \overline{\chi} \to \tau \overline{\tau}$, $\chi \overline{\chi} \to W^+W^-$, etc.). The expected distributions of secondary particles are generated by detailed simulation codes.

The Sun is a very interesting target for such searches, as it is expected to ``collect'' WIMPs in its core while sweeping through the galactic halo. The capture rate $C_c$ is governed by the WIMP-nucleon scattering cross section while the annihilation $C_a$, leading to the observable signature, is described by the annihilation cross section. Neglecting evaporation losses, the number of WIMPs $N_\chi$ in the Sun is given by
  $$\frac{dN_\chi}{dt}=C_c - C_a N_\chi^2\textnormal{.} $$
Assuming that the relevant time scale $\tau = (C_c C_a)^{-1/2}$ is short compared to the age of the Sun, which means that capture and annihilation rate are in equilibrium ($dN_\chi/dt=0$), leads to an annihilation rate 
  $$\Gamma_a = \frac{1}{2} C_a N_\chi^2 = \frac{1}{2} C_c\textnormal{,}$$
which only depends on the scattering cross section. Results from the indirect detection of signals from WIMP annihilation in the Sun can hence be directly compared to the results from underground experiments.

\begin{figure}[t!]
 \centering
 \includegraphics[width=0.5\textwidth]{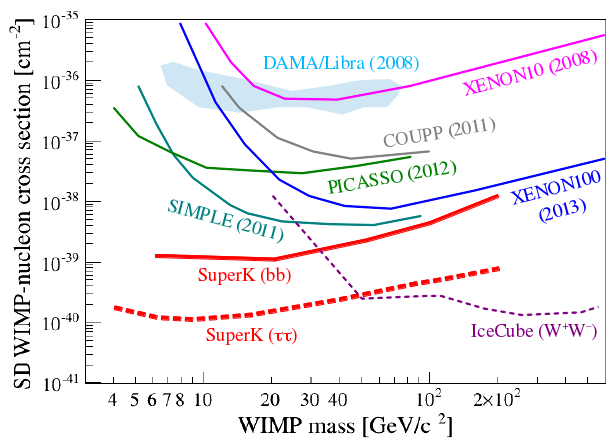}
 \caption{First results on the spin-dependent scattering cross sections for WIMP masses below 10\,GeV/$c^2$ from Super-Kamiokande (red lines)	. Due to the low threshold, the limits are much stronger than other indirect searches (e.g.~from IceCube~\cite{ref::icecube_sd}) and are also superior to direct searches (proton-only couplings). These results are challenging the low-mass WIMP interpretation of the DAMA modulation signal. Fig.~drawn using input presented in~\cite{ref::choi}.}
 \label{fig::superk}
\end{figure}

Using the equilibrium assumption, a recent analysis from Super-Kamiokande places the first constraints from indirect searches on spin-dependent scattering cross sections below $m_\chi=10$\,GeV/$c^2$~\cite{ref::choi}. Super-Kamiokande, located in the Kamioka mine in Japan, is the world's largest water \u{C}erenkov detector with a total mass of 50\,kt (22.5\,kt fiducial target) read out by $\sim$11\,000 photomultipliers. A Monte Carlo study shows that events from low-mass WIMP annihilation to $b\overline{b}$ and $\tau\overline{\tau}$ mainly lead to fully-contained events, where the primary neutrino vertex and the secondary muon track are both contained in the fiducial volume, and which leads to a very clean analysis. No events have been observed above background, excluding spin-dependent WIMP-nucleon cross sections above 10$^{-39..-40}$\,cm$^2$, see Fig.~\ref{fig::superk}. As the Sun mainly consists of protons, these are proton-only couplings.

There are similar results on WIMP annihilations in the Sun from other neutrino detectors, such as from IceCube at South Pole~\cite{ref::icecube_sd} (see also Fig.~\ref{fig::sd}) and ANTARES in the Mediterrean Sea~\cite{ref::hernandez}: Both experiments do not see a signal. In particular ANTARES, which is located in the northern hemisphere, mainly focuses on dark matter annihilation signals from the Galactic Center: in this case one expects upwards going signal muons whereas the main background comes from above.

IceCube is a gigantic neutrino detector installed in the antarctic ice at the South Pole. The 1\,km$^3$ detector corresponds to an instrumented volume of $\sim$1\,Gt. Since the detector is so large, IceCube can search for WIMP annihilation in the Galactic Center, even though it is always above the horizon. Using the outer parts of IceCube as a veto for the inner part, the so-called DeepCore, opens up the possibility to identify neutrino vertices in the DeepCore. These events correspond to a muon track starting in the DeepCore with no track in the veto~\cite{ref::flis}. Preliminary studies show that the DeepCore is sensitive to velocity-averaged annihilation cross sections of $\langle \sigma_A v \rangle \approx 10^{-21..-22}$\,cm$^3$\,s$^{-1}$ for WIMP masses as low as 30\,GeV/$c^2$~\cite{ref::icecube_icrc,ref::wolf}. 

Using neutrinos to search for dark matter has the advantage that their direction is pointing back to the source, as neutrinos are not affected by interstellar magnetic fields etc. However, their low interaction rate and the limited angular resolution of the huge neutrino detectors poses some difficulties. The ``smoking gun'' signature for WIMP annihilation would be a line in the $\gamma$-spectrum of objects expected to have an high WIMP density, e.g., the Galactic Center or dwarf spheroidal galaxies. The expected $\gamma$-flux $\Phi_\gamma$ from dark matter annihilation coming from a solid angle element $\Delta \Omega$ around the coordinates $(\phi,\theta)$ on the sky is given by 
\begin{eqnarray*}\frac{d\Phi_\gamma(E_\gamma, \Delta\Omega(\phi,\theta))}{dE_\gamma} &=&\frac{1}{4 \pi} \frac{\langle \sigma_A v \rangle}{2m_\chi^2} \sum_f \frac{dN_\gamma^f}{dE_\gamma} B_f \\ && \times \int_{\Delta \Omega} d\Omega' \int_{l.o.s.} \rho^2(r(s)) \ ds \textnormal{.}
\end{eqnarray*}
The first part of the equation covers the particle physics of the problem, considering that $\gamma$-particles can be produced with different spectra $dN/dE_\gamma$ in the decay of different primary particles $f$, which are produced in the WIMP annihilation reaction with a certain branching ratio $B_f$. The second part describes the dark matter distribution in Space, by integrating over the dark matter density $\rho(r)$ at distance $r$ along the line of sight $s$.

As WIMPs do not couple directly to photons, the processes $\chi \overline{\chi} \to \gamma \gamma$ or $\chi \overline{\chi} \to \gamma Z$ do not happen at tree-level but only as largely suppressed second-order processes. Still, the two monoenergetic gammas would produce a sharp, distinct spectral feature at $m_\chi$, accompanied by a somewhat broader and lower peak from $\chi \overline{\chi} \to \gamma Z$ at reduced energies.

The Large Area Telescope (LAT) on the Fermi satellite observed the $\gamma$-sky from 20\,MeV to 300\,GeV with a huge field of view. 
In 2012, there has been a claim that a $\gamma$-line at $E_\gamma=130$\,GeV has been found in the LAT data at the Galactic Center~\cite{ref::fermiline}.  An updated analysis of~\cite{ref::meng} uses optimized search regions around the Galactic Center for different dark matter halo models and finds excesses in the spectrum at 130\,GeV and -- less pronounced -- at 110\,GeV. The data is best described by a model assuming an Einasto dark matter profile with a slight offset, leading to a global significance of 5.1\,$\sigma$. However, a similar analysis has been carried out by the Fermi Collaboration~\cite{ref::fermi,ref::mcenery} as well, after reprocessing the data in order to update the energy scale and with newly added data. The global significance of a feature at 130\,GeV in this analysis is only 1.6\,$\sigma$ and seems to be too wide for the instrument's resolution. At this point, it seems clear that more data is needed to clarify the situation; a change in the observation strategy of Fermi could help to acquire this data faster.

\begin{figure}[t]
 \centering
 \includegraphics[width=0.5\textwidth]{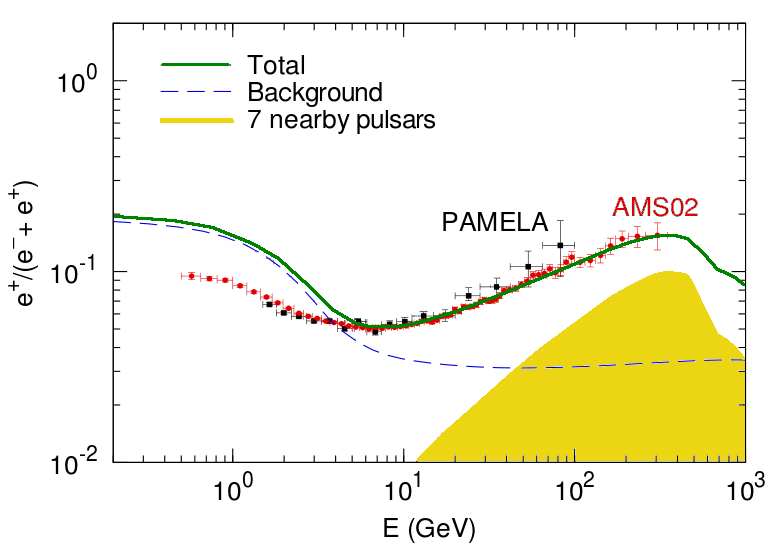}
 \caption{The rising positron fraction $e^+/(e^+ + e^-)$ at higher energies was first observed by PAMELA~\cite{ref::pamela} (black data points) and recently confirmed by AMS-02~\cite{ref::ams2} with high precision (red data points). It can either be explained by dark matter annihilation, or by positrons injected by one or more nearby pulsars. The yellow contribution in the Figure models the summed contribution of seven nearby pulsars, which describes the observed spectral shape very well~\cite{ref::pulsars}. Fig.~adapted from~\cite{ref::pulsars}. }
 \label{fig::pulsars}
\end{figure}

An excess of antiparticles, in particular positrons, is widely discussed as a promising signature for dark matter annihilation since the PAMELA experiment reported a rising fraction of positrons in the total $(e^-+e^+)$-flux for energies above $\sim$5\,GeV~\cite{ref::pamela,ref::boezio}, see Fig.~\ref{fig::pulsars}. At these energies, the astrophysical backgrounds become sub-dominant and the spectrum cannot be simply explained by a modification of the model describing the propagation of the charged cosmic rays. Moreover, the AMS-02 instrument on the International Space Station (ISS) recently confirmed the excess, which keeps increasing up to energies of $\sim$300\,GeV~\cite{ref::ams2,ref::ting}, the current high-energy limit of the instrument. An interpretation of the rising positron fraction in terms of dark matter is possible, however, it conflicts with the non-observation of a similar excess in anti-proton data by PAMELA. The antiproton ratio is in agreement with the expectations from secondary production, which means that dark matter would need to be 'leptophilic' in order to explain the positron excess. The size of the positron signal is also about 3\,orders of magnitude larger than expected from the WIMP miracle, the freeze-out of WIMP dark matter in the expanding Universe. This means that large boost factors are required to generate the observed signature. Possible sources of the boost could be a local over-density of dark matter in nearby sub-halos, or a rate enhancement due to the Sommerfeld effect~\cite{ref::sommerfeld}. An explanation involving nearby pulsars, i.e.~known astrophysical objects which describe the observed data equially well, might be more appropriate one at the moment, see Fig.~\ref{fig::pulsars}~\cite{ref::pulsars}. 

Several difficulties with the interpretation of the positron excess observed by PAMELA and AMS-02 have been pointed out: If WIMPs are Majorana fermions, then two-body leptonic final states of the form $\chi \chi \to \ell \ell$ are $s$-wave suppressed. A three-body final state $\chi \chi \to \ell \ell X$, including a gauge boson $X=W/Z/(\gamma)$ from virtual internal Bremsstrahlung which can carry away the spin, is therefore preferred. The non-observation of an anti-proton excess is again incompatible with such a Majorana annihilation~\cite{ref::weiler1, ref::weiler2}. 

Dwarf spheroidal (dSph) galaxies are small companions of the Milky way which exhibit a very low luminosity. In fact, their dynamics cannot be explained at all by the visible stellar mass, therefore these objects are generally considered as being almost entirely dominated by dark matter. This makes these ${\cal O}(20)$ known objects very interesting targets for indirect WIMP searches as the backgrounds are expected to be rather subdominant. Fermi, for example, has studied the $\gamma$-flux from 10\,dwarf spheroidals in a joint likelihood analysis~\cite{ref::fermidwarfs,ref::mcenery}. No signal has been found, placing significant constraints on WIMP annihilation, in particular at masses below 20\,GeV/$c^2$, where the 90\% CL upper limit excludes the thermal cross section of $\langle \sigma_A v \rangle = 3 \times 10^{-26}$\,cm$^3$\,s$^{-1}$ in the $\chi \chi \to b\overline{b}$ and the $\chi \chi \to \tau^+\tau^-$ channels. 

Segue-1 is a prominent dSph galaxy with a luminosity of only $\sim$350\,solar luminosities $L_\odot$ but a mass-to-light ratio of 3500, pointing towards a large fraction of dark matter. Therefore it has been studied with several instruments such as the air-\u{C}erenkov telescopes (IACTs) VERITAS~\cite{ref::zitzer} and MAGIC~\cite{ref::aleksic}. Their analyses also yield null-results and are in agreement with the Fermi limits from dSphs mentioned above. The high-altitude water-\u{C}erenkov observatory HAWC complements the results from Fermi and IACTs by extending the limits up to WIMP masses of 100\,TeV/$c^2$~\cite{ref::baughman}. It has been pointed out that a large number of low-mass sub-halos, without any stars and light, are prediced by simulations of the Universe, such as Aquarius~A~\cite{ref::aquarius} and Via~Lactea~II~\cite{ref::vialactea}. However, many of these halos are below the mass resolution of the simulations. A new method exploring this mass range suggests that many extended sub-halos should exist which could be promising candidates for indirect searches~\cite{ref::klein}.

Instead on focusing on the annihilation signature from astronomical objects, e.g., the Sun, the Galactic Center, dSph galaxies, etc., one can also use the entire Universe to constrain dark matter annihilation in a calorimetric approach~\cite{ref::galli, ref::iocco}. The decay products from the annihilation interact with the cosmological environment which is being heated and ionized. This process is most effective in the early Universe, when the densities of the gas and the WIMPs were high, and one should expect an impact on the cosmic microwave background (CMB) spectrum. The additional particles from the annihilation process would mainly impact the CMB multipole spectrum around $\ell\sim1000$, a feature which is not observed by the measurements. This leads again to tight constraints on WIMP annihilations, in particular on low-mass WIMPs of $m_\chi \le 10$\,GeV/$c^2$ as it reaches below the thermal cross section for the $\chi\chi \to e^+e^-$ channel.

\section{Collider Searches}\label{sec::collider}

The last approach which we will discuss in this article is the production of WIMP dark matter at colliders, in particular at a hadron collider such as the LHC~\cite{ref::colliderreview}. Being only very weakly interacting, WIMPs do not deposit any energy in the detectors (similar to neutrinos), and one needs to look at the total energy and momentum budget of an event, as measured in all detector components, in order to indentify such particles via a missing energy signal $E_\textnormal{\tiny miss}$. In $pp$-collisions at the LHC, the initial longitudial momentum of the partons is unknown, hence one can only use the missing energy in the transversal plane, $E_\textnormal{\tiny miss}^T$ for the WIMP search. 

The two general-purpose detectors at LHC, ATLAS~\cite{ref::atlas} and CMS~\cite{ref::cms}, both provide almost $4\pi$ coverage around the interaction point and were designed to (successfully) search for the Higgs particle~\cite{ref::higgs}, for new physics, as well as for precision tests of the Standard Model. The experiments measure the momentum of charged particles, the energy of electromagnetic showers caused by electrons and gammas, and the energy of hadronic showers from strongly-interacting particles, combining this information to search for new phenomena such as dark matter~\cite{ref::pinfold}. Astrophysical uncertainties are completely absent in collider results, however, the very limited time a particle spends in the detector will make it almost impossible to proof from collider data alone, that a detected candidate is the dark matter particle.

\begin{figure}[t]
 \centering
 \includegraphics[width=0.35\textwidth]{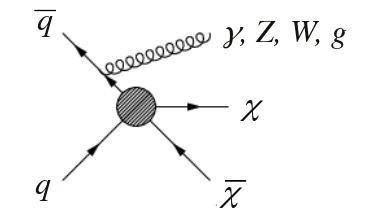}
 \caption{The most generic approach to search for WIMPs at a hadron collider is to focus on events which could be due to WIMP pair-production together with initial (or final) state radiation of a $\gamma$, $Z$, $W$, or a gluon.}
 \label{fig::collider}
\end{figure}

As pair-production of WIMPs of the type 
  $$q\overline{q} \to \chi \overline{\chi}$$ 
is invisible to the detectors, as absolutely nothing would be seen by the various detector components, the most generic approach to search for WIMP production at a hadron collider is to search for pair-production associated with initial (or final) state radiation
  $$q\overline{q} \to \chi \overline{\chi} + X\textnormal{,}$$
with $X$ being a gamma, $Z$- or $W$-boson, or a gluon, see Fig.~\ref{fig::collider}. The unknown coupling of WIMPs $\chi$ to standard model fermions $q$ can be described in a largely model-independent fashion using effective field theories and contact operators~\cite{ref::effield}. Depending on the choice of the operators, the interaction is similar to direct (spin-independent, spin-dependent) or indirect searches ($s$-wave and $p$-wave annihilation). The initial state radiation leads to an imbalance in the detected energy and momentum, and the WIMP search is based on events with a high $E_\textnormal{\tiny miss}^T$ plus a single particle track or jet. The searches are therefore also referred to as monophoton, mono-$Z$, mono-$W$ and monojet searches.

\begin{figure}[b!]
 \centering
  \includegraphics[width=0.48\textwidth]{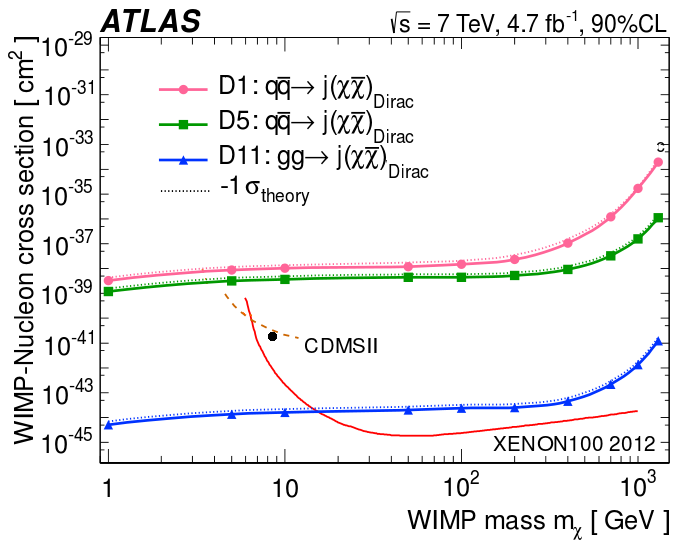}
 \caption{Constraints on the spin-independent (Dirac fermion) WIMP-nucleon scattering cross section from a monojet search of ATLAS~\cite{ref::atlas_monojet}. No excess of events was found above the background expectation and the lines labeled with D1, D5, D11 correspond to 90\% CL upper limits for different effective interaction operators. For comparison, the direct detection limits from CMDS-II (low mass analysis~\cite{ref::cmdslow}) and XENON100~\cite{ref::xe100} are also shown, as well as the point with the highest likelihood from the CDMS-Si signal claim~\cite{ref::cdms_si} (black point). Figure adapted from~\cite{ref::atlas_monojet}.}
 \label{fig::lhc}
\end{figure}

An example of a result from a monojet search with ATLAS is shown in Fig.~\ref{fig::lhc}, using LHC data with a center-of-mass energy of $\sqrt{s}=7$\,TeV and an integrated luminosity of 4.7\,fb$^{-1}$~\cite{ref::atlas_monojet}. The event selection criteria usually require $E_\textnormal{\tiny miss}^T \approx 120 \ldots 500$\,GeV, a well-reconstructed jet with a transverse momentum $p_T>110$\,GeV, and no additional lepton or jet. $Z$-boson production together with a jet, with the $Z$ decaying into two neutrinos, is the main Standard Model background for this search, which is determined by data-driven methods. As no excess of events above the Standard Model expectation has been been found, the experiment could set upper limits on the WIMP-nucleon scattering cross section by considering the right model for the interaction in Fig.~\ref{fig::collider}. Besides the choice of a cut-off scale $\Lambda$, the results are model independent in the sense that the only assumption entering the analysis is that no other particles can be directly produced in the $pp$-collision. 

The result can now be directly compared to results from direct detection searches such as XENON100~\cite{ref::xe100}. One can immediately see that the sensitivity of the collider searches does not bend upwards at low WIMP masses, placing hard constraints in the ${\cal O}(10)$\,GeV/$c^2$ region, where several direct searches observe an excess of events (see Sect.~\ref{sec::direct}). However, the limits get significantly weaker if other effective interaction operators are assumed in the analysis. A similar result is being obtained for spin-dependent interactions. The analyses based on monophotons and mono-bosons yields slightly weaker but in general similar constraints and there is also no significant difference between the results from ATLAS and CMS.

Another interesting constraint on WIMPs comes from the recent detection of the Higgs particle at $m_H = 126$\,GeV/$c^2$~\cite{ref::higgs}. In many models (also if the Higgs is the Standard Model Higgs), the Higgs particle will couple strongly to the dark matter particle if its mass is lighter than $m_H/2=63$\,GeV/$c^2$~\cite{ref::higgstodm}. Since Higgs decays into WIMPs would not be detected, one tries to determine the branching ratio $BR(H \to \textnormal{invisible})$, which was found to be $\le$0.2 at 95\% CL for a Higgs with Standard Model couplings by ATLAS and CMS. For Majorana fermion dark matter, this would lead to limits on spin-independent WIMP-nucleon interactions, mediated by Higgs exchange, which are better than the one of XENON100 for $m_\chi \lesssim 55$\,GeV/$c^2$, again challenging the dark matter interpretation of the excesses seen by DAMA, CRESST-II, CoGeNT and CDMS-Si (see Sect.~\ref{sec::direct}). The constraints are lowered by a factor~2 if the dark matter particle is a Dirac fermion.

\section{Exotic Dark Matter}

In this review, we use the phrase ``exotic'' for all dark matter models which do not predict WIMPs. There is a plethora of theories and non-WIMP dark matter candidates, however, in this article we will only mention two selected possibilities which were presented at ICRC 2013.

The axion is a candidate which is very well motivated by the strong CP problem but exhibits features which make it very different from WIMPs~\cite{ref::axion,ref::nigelsmith}. Dark matter axions are expected to have masses in the few $\mu$eV range. As dark matter needs to be non-relativistic (``cold''), their low mass requires that dark matter axions have been created non-thermally but in a phase transition (vacuum realignment). The parameter space relevant for the dark matter problem is probed by microwave cavity experiments such as  ADMX~\cite{ref::admx}, which search for signals of axion-microwave photon conversion in a strong magnetic field. Axion-like particles (ALPs) and heavy photons are generalizations of the dark matter axions to other mass ranges and couplings. There are several experiments searching for ALPs, examples are XMASS, which recently published a new limit on the coupling of ALPs to electrons~\cite{ref::yamashita,ref::xmass_axions}, and H.E.S.S., which placed limits on ALPs by observing the active galaxy PKS 2155-304~\cite{ref::brun}.

Primordial black holes (BPH) produced in the big bang are another possibility of a dark matter candidate. A new study~\cite{ref::barnacka} constrains PBHs using femto-lensing methods, where ``femto'' refers to the very small angular distance between the gravitationally lensed images. One expects that the spectra from $\gamma$-ray bursts detected by the Fermi satellite show an interference pattern due to femtolensing on PBHs, since the photon wavelength is comparable to the Schwarzschild radius of the lens. No such deviations have been detected and new limits on the abundance of PBHs in the mass region of $10^{16..18}$\,g could be derived.

\section{Conclusion}

As of summer 2013, when it was discussed at ICRC 2013 in various aspects, the global picture regarding the detection of dark matter remains rather ambiguous. Event excesses (or ``anomalies''~\cite{ref::weiner}) observed in direct and indirect detection experiments, which could be explained by dark matter scattering or annihilation, are confronted with several null-results from direct, indirect and collider searches. 

The hints for WIMP scattering in underground detectors appear to cluster at rather low WIMP masses around $m_\chi \approx 10$\,GeV/$c^2$, however, the published evidence regions are not all mutually consistent. There is tension with the latest result from XENON100~\cite{ref::xe100} which for example clearly excludes the point of highest likelihood in the CDMS-Si region~\cite{ref::cdms_si,ref::xe100_mcmatch}. At lowest masses, however, there is some parameter space remaining where the two results can be compatible~\cite{ref::frandsen}. The null-results from the LHC (see Sect.~\ref{sec::collider}) and indirect searches (see Sect.~\ref{sec::indirect}, in particular the new limit from Super Kamiokande~\cite{ref::choi}), and various other studies (e.g.~\cite{ref::galli,ref::iocco}) further increase the tension. 

In general, a dark matter explanation of the excesses seen in indirect searches, mainly the possible $\gamma$-line from the Galactic Center and the rising positron fraction, requires a more massive WIMP ($m_\chi>100$\,GeV/$c^2$) than the excesses seen in direct detection ($m_\chi\approx10$\,GeV/$c^2$). However, another indication from indirect detection, a residual $\gamma$-emission around the Galactic Center, could be explained by dark matter annihilation of particles of lower mass (7-45\,GeV/$c^2$)~\cite{ref::linden}. From the collider side, there is absolutely no indication of a signal to-date. A full review of the various excesses and their possible dark matter origin has been presented in~\cite{ref::weiner}.

We conclude that as of summer 2013, the nature of the dark matter particle remains a mystery and that search for dark matter is still ongoing. Impressive progress has been made at direct, indirect and collider searches, and the next generation datasets, analyses and instruments will provide valuable inputs to answer the question: {\it What is dark matter?}

\vspace*{0.8cm}
\footnotesize{{\bf Acknowledgment:}{ I would like to thank the organizers of ICRC 2013 for inviting me to this great conference.}}

\end{document}